\documentclass[12pt]{article}
\usepackage{graphicx}
\usepackage{subfigure}
\usepackage{amsmath}
\usepackage{amssymb}

\begin{document}

\begin{center}
\begin{large}
\title\\{ \textbf{An Improved analysis of Isgur-Wise Function for Heavy Light Mesons in $D$ dimensional Potential Model.}}\\\

\end{large}

\author\

\textbf{$Sabyasachi\;Roy^{\emph{1}}\footnotemark\:and\:D.\:K.\:Choudhury^{\emph{1,2}}$ } \\\
\footnotetext{Corresponding author (On leave from Karimganj College,Assam, India). e-mail :  \emph{sroy.phys@gmail.com}}
\textbf{1}. Department of Physics, Gauhati University, Guwahati-781014, India.\\
\textbf{2}. Physics Academy of The North East, Guwahati-781014, India.

\begin{abstract}
This letter reports some improvement of the formalism and analysis of our very recent work \cite{SR1} on the studies of Isgur-Wise function for heavy-light mesons following higher dimensional string inspired potential model.
\end{abstract}
\end{center}


\section{Introduction:}
Very recently we have reported our studies on heavy-light mesons and its Isgur-Wise function(IWF) in higher dimension following Nambu-Goto string inspired higher dimensional potential model \cite{SR1}, taking its Luscher term as parent in Solving higher dimensional Schrodinger equation through perturbation method. In that article, while developing meson wave function and subsequently in the studies of  IWF and its derivatives (slope and curvature), we have carried out all the integrations (equations (15) to (34) of ref. \cite{SR1}) in three dimensional space taking surface element to be $4\pi r^2 dr $. Very recently, we have reported similar studies on heavy light mesons and its IWF considering linear confinement term of the said potential to be parent \cite{SR2}. There we have introduced higher dimensional surface element $D C_D r^{D-1}dr$ \cite{Eduarda} in place of $4\pi r^2 dr $ while carrying out all the integrations involved in the studies of normalisation constant, energy terms, IWF and its derivatives in D-dimensional space. \\
This letter reports improvement of the formalism and numerical calculations of ref. \cite{SR1}(Luscher parent case) considering mesons to be in D-dimensional space (as in ref. \cite{SR2}). Our improved results though give back the corresponding results in three dimensional QCD (ref. \cite{NSB}) but still they are higher in range than the expectations, as in the previous version of the work (ref.\cite{SR1}). Here, following convergence condition of the perturbation series, we further introduce cut-off to upper limits of integrations in calculations of IWF and its derivatives, which is dimension dependent. Results with this cut-off are now found to be within the range of standard expectations.\\
A gist of the improvement in formalism is reported in $\S 2$ , while results and our conclusion are reported in $\S 3$.
\section{Formalism:}
The unperturbed wave function $\Psi^{0}(r)$ remains the same as in equation (14) of ref \cite{SR1}.
\begin{equation}
\Psi^{0}(r)= N r^{\frac{D-3}{2}}e^{-\mu\gamma r}
\end{equation}
The normalisation condition in D spatial dimension is:
\begin{equation}
\int_0^{\infty} D. C_D. r^{D-1} \mid \Psi(r)\mid ^{2}dr =1
\end{equation}
where $C_D=\frac{\pi^{D/2}}{\Gamma(\frac{D}{2}+1)}$ \cite{Eduarda}. From this we obtain $N$ to be:
\begin{equation}
N=\frac{1}{(D. C_D)^{1/2}}[\frac{(2\mu\gamma)^{2D-3}}{\Gamma (2D-3)}]^{1/2}
\end{equation}
The eigenenergy $E$ now comes out as:
\begin{equation}
E=W^{0}=-\int_0^{\infty}\frac{\gamma}{r} D. C_D r^{D-1}\mid \Psi^{0}(r)\mid^{2}dr =-\frac{\mu\gamma^{2}}{D-2}
\end{equation}
Next, while calculating the perturbed part of the wave function (as in \S2.3 of ref \cite{SR1}), we find the modified expression of perturbed energy eigenvalue as:
\begin{equation}
W^{\prime}=\int_0^{\infty} D.C_D r^{D-1}H^{\prime}\mid \Psi^{0}(r)\mid^{2}dr =\sigma\frac{2D-3}{2\mu\gamma}+\mu_0
\end{equation}
This we use in applying the Dalgarno's method of perturbation to extract the perturbed eigenfunction, which is now:
\begin{equation}
\Psi^{\prime}(r)=  -\frac{\sigma (2D-3)}{6\gamma}r^{2}r^{\frac{D-3}{2}}e^{-\mu\gamma r}
\end{equation}
Our modified total wave function is:
\begin{equation}
\Psi^{tot}(r)=\Psi^{0}(r)+\Psi^{\prime}(r)=N_1(1-k r^2)r^{\frac{D-3}{2}}e^{-\mu\gamma r}
\end{equation}
Here, $k=\frac{\sigma (2D-3)}{6\gamma}$.
Employing equation (2), the normalisation $N_1$ for total wave function $\Psi^{tot}(r)$ can now be expressed as:
\begin{equation}
N_1=\frac{1}{(D.C_D)^{1/2}}.\frac{1}{[\frac{\Gamma(2D-3)}{(2\mu\gamma)^{2D-3}}-2.k.\frac{\Gamma(2D-1)}{(2\mu\gamma)^{2D-1}}+k^2.\frac{\Gamma(2D+1)}{(2\mu\gamma)^{2D+1}}]^{1/2}}
\end{equation}

Here, we take a note of the following points.
\begin{itemize}
  \item At $D=3$, our present expressions for $\Psi^{0}(r),\Psi^{tot}(r),N, N_1$ gives back exactly the corresponding expressions of ref. \cite{NSB}.
  \item With only Luscher term in potential ($\sigma=0$) normalisation constant of equation (8) becomes:
  \begin{equation}
N_1=\frac{1}{(D. C_D)^{1/2}}[\frac{(2\mu\gamma)^{2D-3}}{\Gamma (2D-3)}]^{1/2}
\end{equation}
This is our normalisation constant N for unperturbed wave function $ \Psi^{0}(r)$ as in eqn. (3).
\end{itemize}
This shows that the three dimensional correspondence is preserved in this approach also, as in ref \cite{SR1}. \\
Now, the Taylor series expansion of IWF around zero recoil point $y=1$ will remain the same as in equation (28) of ref \cite{SR1}:
\begin{equation}
\xi(y)=1-\rho^2 (y-1) +C(y-1)^2+\cdots\cdots
\end{equation}
By equation (33) of ref. \cite{SR1}, the improved formulation of IWF in terms of meson wave function in higher spatial dimension is:
\begin{equation}
\xi(y)=\int_0 ^\infty D C_D r^{D-1} |\Psi(r)|^2\cos(pr)dr
\end{equation}
From equations (10) and (11), we obtain expressions for slope and curvature:
\begin{eqnarray}
\rho^2 = D C_D\mu^2\int_0^\infty r^{D+1}|\Psi(r)|^2dr \\
C= \frac{1}{6}D C_D\mu^4\int_0^\infty r^{D+3}|\Psi(r)|^2dr
\end{eqnarray}
Using these general expressions we calculate $\rho^2$ and $C$ considering both the unperturbed and total wave functions. \\
\textbf{With $\Psi^{0}(r)$:}
\begin{eqnarray}
\rho^2=\frac{(2D-2)(2D-3)}{4\gamma^{2}} \\\
C=\frac{2D(2D-1)(2D-2)(2D-3)}{96\gamma^{4}}
\end{eqnarray}
\textbf{With $\Psi^{tot}(r)$:}
\begin{eqnarray}
\rho^{2}=\mu^{2}\frac{[\frac{\Gamma(2D-1)}{(2\mu\gamma)^{2D-1}}-2k\frac{\Gamma(2D+1)}{(2\mu\gamma)^{2D+1}}+k^2\frac{\Gamma(2D+3)}{(2\mu\gamma)^{2D+3}}]}{[\frac{\Gamma(2D-3)}{(2\mu\gamma)^{2D-3}}-2k\frac{\Gamma(2D-1)}{(2\mu\gamma)^{2D-1}}+k^2\frac{\Gamma(2D+1)}{(2\mu\gamma)^{2D+1}}]} \\
C=\frac{\mu^{4}}{6}\frac{[\frac{\Gamma(2D+1)}{(2\mu\gamma)^{2D+1}}-2k\frac{\Gamma(2D+3)}{(2\mu\gamma)^{2D+3}}+k^2\frac{\Gamma(2D+5)}{(2\mu\gamma)^{2D+5}}]}{[\frac{\Gamma(2D-3)}{(2\mu\gamma)^{2D-3}}-2k\frac{\Gamma(2D-1)}{(2\mu\gamma)^{2D-1}}+k^2\frac{\Gamma(2D+1)}{(2\mu\gamma)^{2D+1}}]} \end{eqnarray}
At $D=3$ , equations (16) and (17) simplifies to equations (40) and (41) of ref. \cite{SR1}. \\
Here we find:
\begin{itemize}
  \item In both the cases of wave function consideration, at $D=3$ present expressions for $\rho^2$ and $C$ give the corresponding expressions of ref. \cite{NSB} when $\gamma$ is replaced by $\frac{4\alpha_s}{3}$ and $\sigma$ by $b$ of Cornell potential.
  \item Also, equations (16) and (17) simplifies to equations (14) and (15) respectively, when there is only Luscher term in potential, with no confinement ($\sigma=0$).
  \item Furthermore, although equations (35-36) and (38-39) of ref. \cite{SR1} reflect the same features, still it is very much clear from our present formalism that our equations (14-17) are more generalised expressions in D-dimensional space.
\end{itemize}

\section{Results and Conclusion:}
With this improvement of formalism over that in ref \cite{SR1}, we now study the the variation of IWF and its derivatives with dimension. The input parameters for quark masses are taken from ref. \cite{Vinodkumar}. Results for $\rho^2$ and $C$ for B meson are reported in Table-1 and Figures 1,2.
\begin{table}[!htbp]\scriptsize
\begin{center}
\caption{Slope and Curvature for B meson.} \label{cross}
\begin{tabular}{|c|cc|cc|}
  \hline
 D   &$ with\;\; \Psi^{0}(r)$ &    & $ with\;\;  \Psi^{total}(r)$ &    \\

     & $\rho^2$ & C  & $\rho^2$ & C  \\
  \hline \hline
   3        & 43.7594 & 797.869 & 204.376 & 11183.1   \\    \hline
   5        & 51.0526 & 698.135 & 120.415 & 3331.13   \\    \hline
   10       & 55.1044 & 628.469 & 83.2296 & 1379.3    \\    \hline
   15       & 56.2621 & 607.125 & 73.8462 & 1027.93   \\    \hline
   20       & 56.8105 & 596.822 & 69.5939 & 886.884   \\    \hline
   25       & 57.1303 & 590.757 & 67.1706 & 811.511   \\    \hline
   50       & 57.7505 & 578.892 & 62.5922 & 678.947   \\    \hline
   $\infty$ & 58.3460 & 567.38  & 58.348  &  567.43   \\
   (asym) &&&& \\ \hline
  \hline

\end{tabular}
\end{center}
\end{table}

\begin{figure}[!htbp]
    \centering
    \subfigure[$\rho^{2}$ vs $D$ with $\Psi^{0}(r)$]
    {
        \includegraphics[width=3.0in]{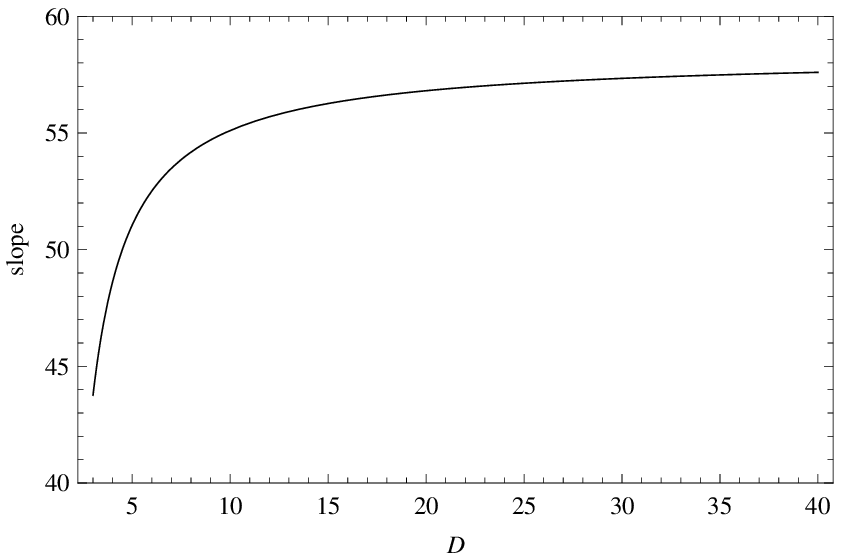}
        \label{fig:first_sub}
    }
        \subfigure[C vs $D$ with $\Psi^{0}(r)$ ]
    {
        \includegraphics[width=3.0in]{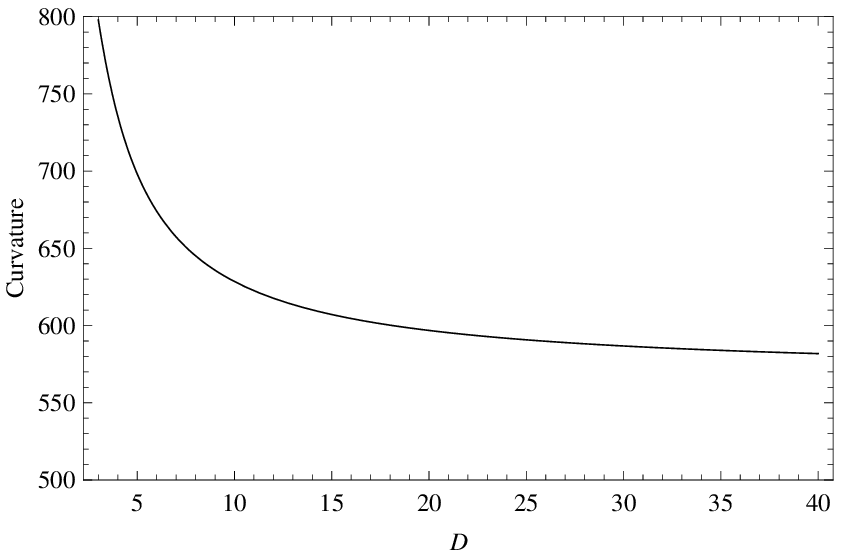}
        \label{fig:second_sub}
    }
\caption{Variation of $\rho^{2}$ and $C$ vs $D$ for B meson: with $\Psi^{0}(r)$ }
\label{fig:sample_subfigures}
\end{figure}

\begin{figure}[!htbp]
    \centering
    \subfigure[$\rho^{2}$ vs $D$ with $\Psi^{tot}(r)$]
    {
        \includegraphics[width=3.0in]{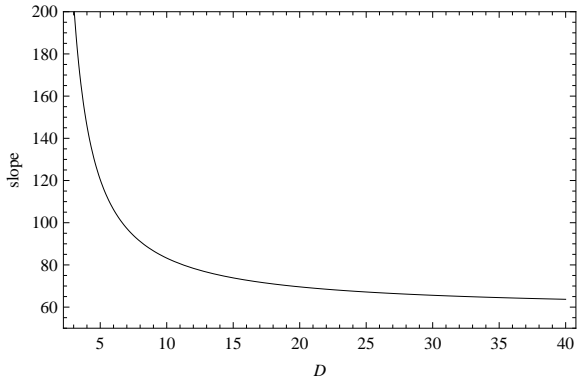}
        \label{fig:first_sub}
    }
        \subfigure[C vs $D$ with $\Psi^{tot}(r)$ ]
    {
        \includegraphics[width=3.0in]{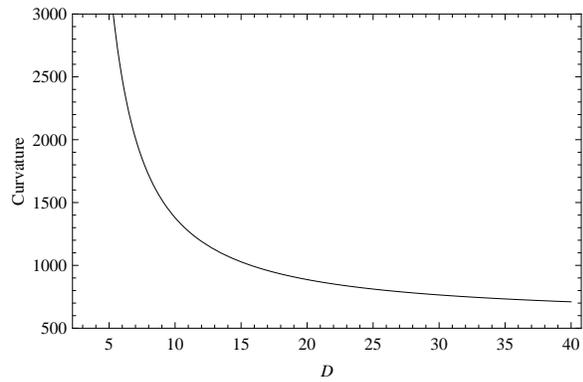}
        \label{fig:second_sub}
    }
\caption{Variation of $\rho^{2}$ and $C$ vs $D$ for B meson: with $\Psi^{tot}(r)$ }
\label{fig:sample_subfigures}
\end{figure}
The variation of $\xi(y)$ with $y$ for B meson at different dimension values are depicted in Figure 3, where zero recoil condition is maintained all through.
\begin{figure}[!htbp]
    \centering
    \subfigure[$\xi(y)$ vs y for diff. $D$ with $\Psi^{0}(r)$]
    {
        \includegraphics[width=3.0in]{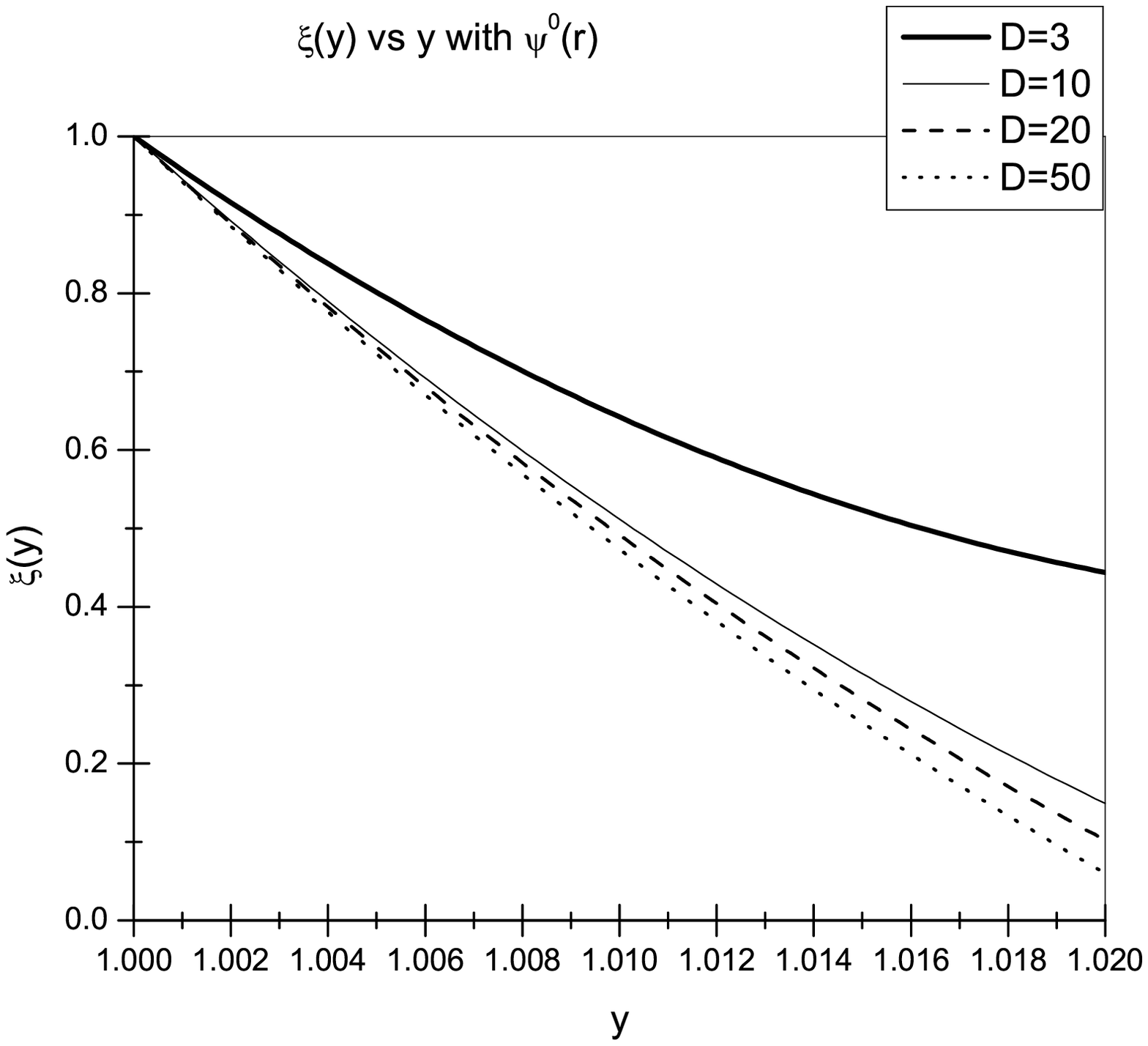}
        \label{fig:first_sub}
    }
        \subfigure[$\xi(y)$ vs y for diff. $D$ with $\Psi^{tot}(r)$ ]
    {
        \includegraphics[width=3.0in]{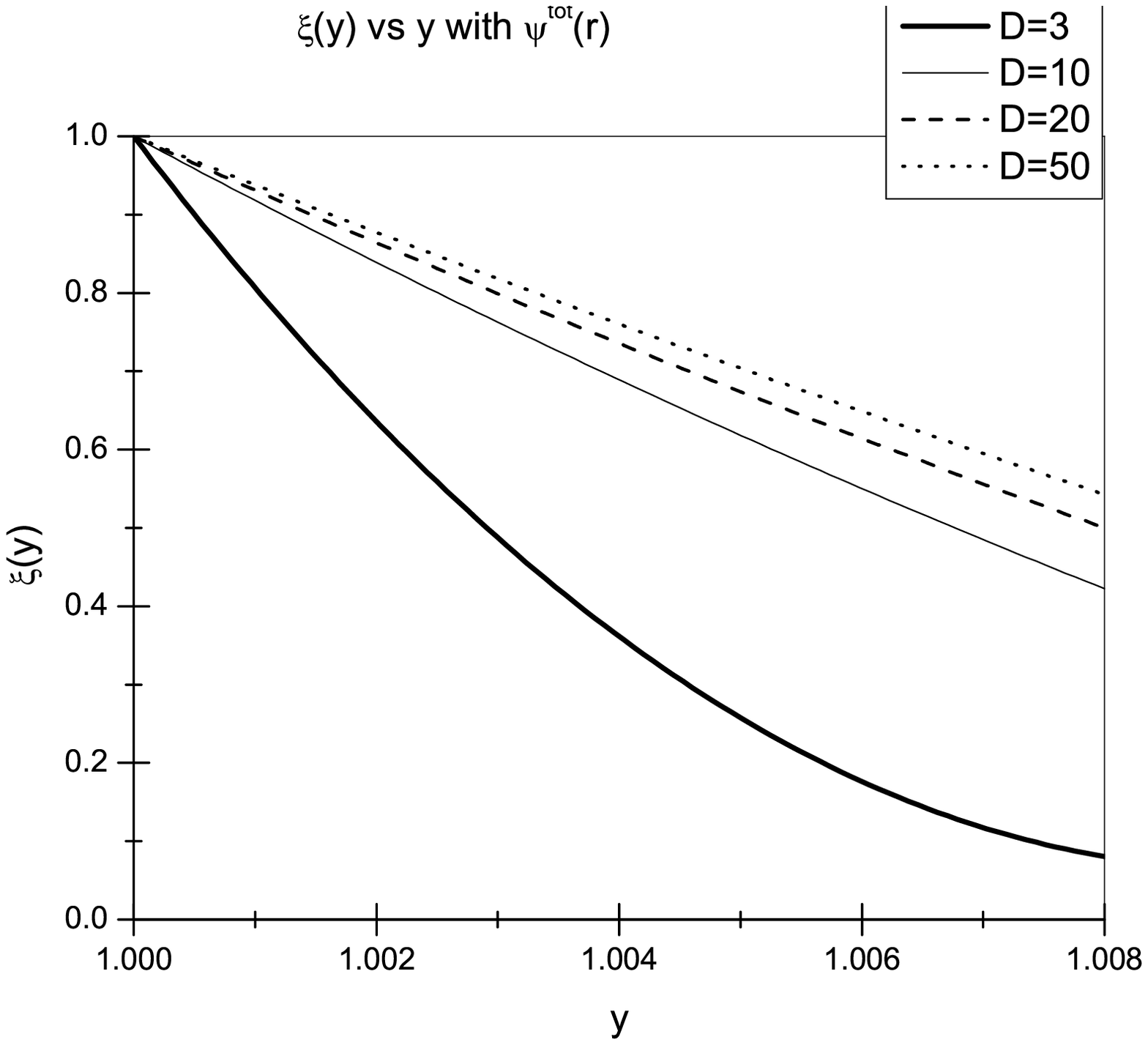}
        \label{fig:second_sub}
    }
\caption{Variation of IWF with y for B meson with diff. $D$ values}
\label{fig:sample_subfigures}
\end{figure}
Our result of Table 1 shows that at $D=3$ our present results for $\rho^{2}$ and $C$ are much lower than in the earlier case(ref. \cite{SR1}). Though $\rho^{2}$ value shows the decreasing trend in the calculation with total wave function (Fig. 2(a)), but with unperturbed wave function, it increases (Fig. 1(a)). The trend of curvature $C$ is decreasing in both the cases of wave function consideration (Fig. 1(b) and 2(b)). However, with both unperturbed and total wave function, $\rho^{2}$ and $C$ values approach the same asymptotic limits which are $\rho^{2}_{asym}\approx 58.34$ and $C_{asym}\approx 567.4$ respectively. \\
The results for slope and curvature of IWF in different theoretical models are already reported in Table 2 of ref. \cite{SR1}. The expected range of slope parameter is 1 to 1.6, whereas curvature parameter value should be less than unity \cite{Yaouanc}. It is evident that our results are in much higher range than the expectation. Here, we take a note of the fact that when we choose the parent and perturbation terms, the underlying assumption is that the perturbative term should not have dominant effect over and above the parent term in potential. In Nambu string inspired higher dimensional potential, the Luscher term is only a non-perturbative correction to the linear confinement term \cite{Bali} and being linearly dependent on dimension parameter, it becomes dominant only at higher dimension. In that view, the consideration of taking Luscher term in parent Hamiltonian is only suitable for higher range of dimension parameter $D$. \\
While looking for the stability of our perturbative approach for whole range of $D$,  we apply the convergence condition of the wave function, as applied in ref \cite{SR2}. This will give:
\begin{eqnarray}
\mid \Psi^{\prime}(r)\mid < \mid \Psi^{0}(r) \mid
\Rightarrow r^2 < \frac{6\gamma}{\sigma(2D-3)}
\end{eqnarray}
This condition gives us the cut-off to the upper limit of integrations involved in $\rho^{2}$ and $C$, which is:
\begin{equation}
r_0 = \sqrt{\frac{6\gamma}{\sigma(2D-3)}}
\end{equation}
We find that unlike as in ref \cite{SR2}, here this upper cut-off limit of integration $r_0$ is independent of the flavour of meson, depending only upon dimension parameter $D$. Table 2 gives the values of this cut-off for different $D$.
\begin{table}[!htbp]\scriptsize
\begin{center}
\caption{Values of $r_0$ for different $D$}
\begin{tabular}{|cc||cc|}
  \hline
  D & $r_0$  & D & $r_0$  \\
   \hline \hline
  3   &  1.7152   &  15   &  1.5127  \\
  4   &  1.6272   &  20   &  1.5054  \\
  5   &  1.5880   &  25   &  1.5011  \\
  10  &  1.5285   &  50   &  1.4931  \\
    \hline
\end{tabular}
\end{center}
\end{table}
With this cut-off, we also calculate slope and curvature of IWF following equations (12-13) and the results for B meson are reported in Table 3.
\begin{table}[!htbp]\scriptsize
\begin{center}
\caption{Slope and Curvature for B meson (calculated with cut-off $r_0$).} \label{cross}
\begin{tabular}{|c|cc|cc|}
  \hline
 D   &$ with\;\; \Psi^{0}(r)$ &    & $ with\;\;  \Psi^{total}(r)$ &    \\

     & $\rho^2$ & C  & $\rho^2$ & C  \\
  \hline \hline
   3        & 0.1880  & 0.0071  & 0.0928  &  0.0020 \\    \hline
   5        & 0.2482  & 0.0108  & 0.1570  &  0.0047 \\    \hline
   10       & 0.2867  & 0.0139  & 0.3010  &  0.0153 \\    \hline
   15       & 0.2983  & 0.0149  & 0.3073  &  0.0158 \\    \hline
   20       & 0.3041  & 0.0155  & 0.3093  &  0.0160 \\    \hline
   25       & 0.3076  & 0.0158  & 0.3110  &  0.0161 \\    \hline
   50       & 0.3148  & 0.0165  & 0.3155  &  0.0166 \\    \hline
   \hline

\end{tabular}
\end{center}
\end{table}
From Table 3, we see that values of slope and curvature increases marginally with increase in $D$, but more importantly these remains below the limit of the expected range. This behaviour is similar to that obtained in ref \cite{SR2} where we have considered linear term as parent. This variation of $\rho^2$ and $C$ for B meson is also clear from Figure 4.
\begin{figure}[!htbp]
    \centering
    \subfigure[$\rho^{2}$ vs $D$]
    {
        \includegraphics[width=3.2in]{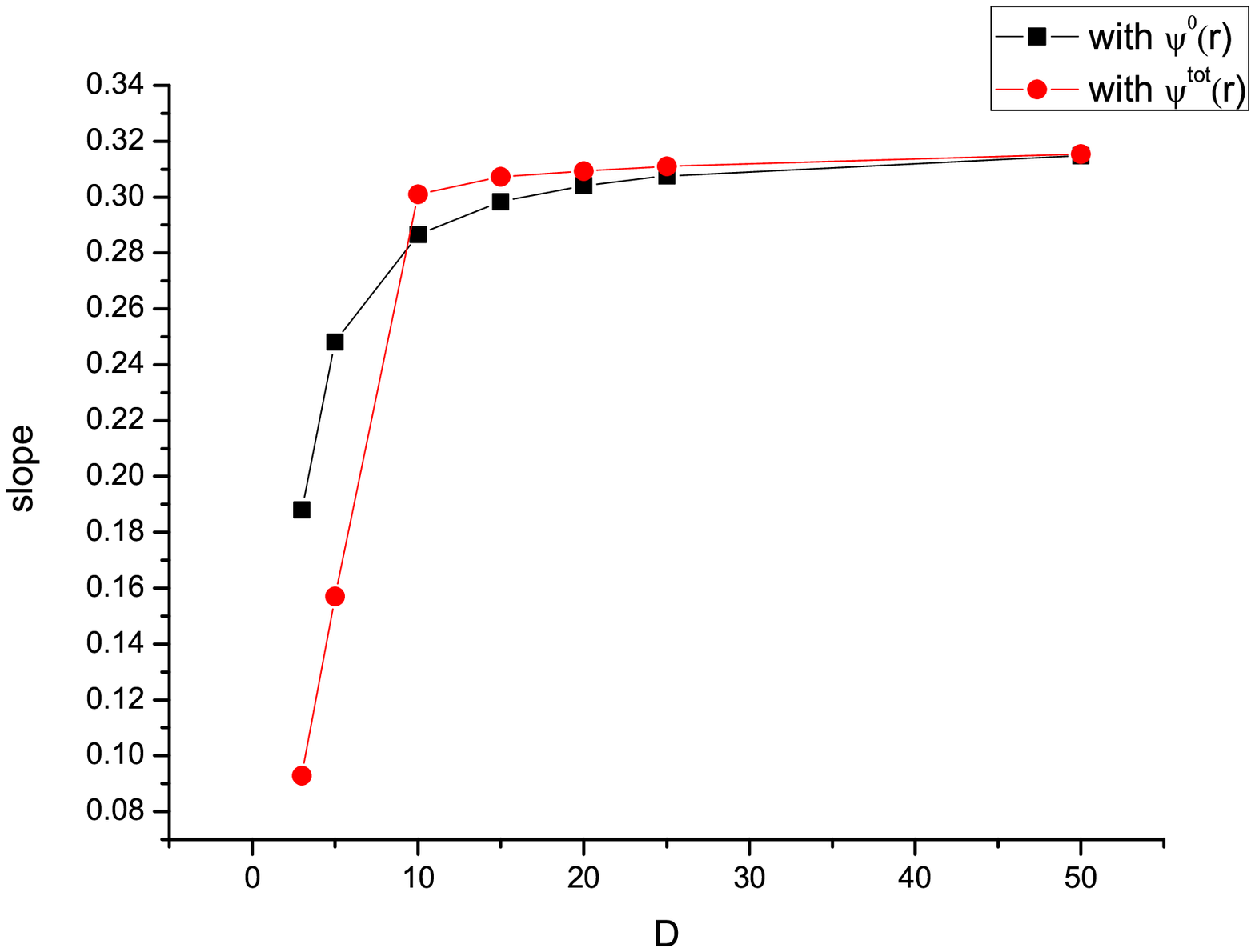}
        \label{fig:first_sub}
    }
        \subfigure[C vs $D$ ]
    {
        \includegraphics[width=3.2in]{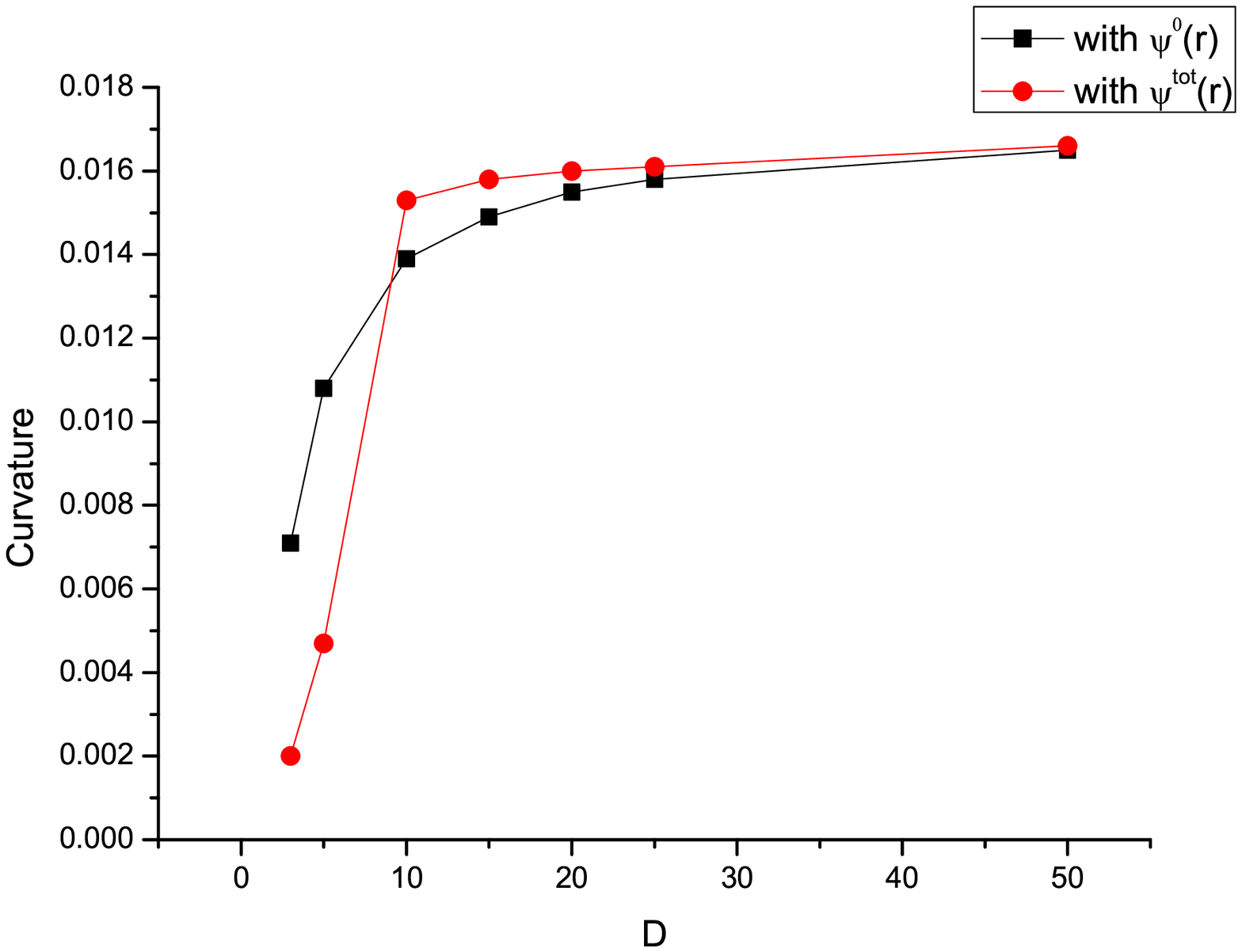}
        \label{fig:second_sub}
    }
\caption{Variation of $\rho^{2}$ and $C$ vs $D$ for B meson. }
\label{fig:sample_subfigures}
\end{figure}
From Figure 4, we find that with increasing $D$, derivatives of IWF increases steadily and ultimately attain upper asymptotic values. For B meson, for both the wave functions, these asymptotic values we find to be $\rho^2_{asym} = 0.3215$ and $C_{asym}=0.0172$. Similar nature of variation of slope and curvature of IWF is also found in other D and B sector mesons. Here, we take a note of the point that, in the present approach, with Luscher term as parent, the range of cut-off (Table 2) is lower than that in ref. \cite{SR2}, resulting in lower range of values of $\rho^{2}$ and $C$. We have worked with fixed value of string constant ($\sigma=0.178 GeV^2$). This  $\sigma$ value may also be varied under different string pictures and accordingly our results may also change(improve). But, independent of lower range of our results or otherwise, in this improved analysis we have deduced a more generalised formulation of $\rho^{2}$ and $C$ taking into care stability of the perturbative approach.

\end{document}